\documentclass[aps,prl,twocolumn,superscriptaddress]{revtex4-1}
\usepackage{graphicx}
\usepackage[rgb,dvipsnames]{xcolor}
\usepackage{amsmath}
\usepackage{amsthm, amssymb}
\usepackage{epsfig}
\usepackage{hyperref}
\usepackage{bm}

\usepackage{natbib}

\usepackage[normalem]{ulem}

\usepackage{placeins}
\usepackage[symbol]{footmisc}

\usepackage{tikz}
\usetikzlibrary{trees}

\begin{document}

\title{Contact-mediated signaling enables disorder-driven transitions in cellular assemblies}

\author{Chandrashekar Kuyyamudi}
\affiliation{The Institute of Mathematical Sciences, CIT Campus, Taramani, Chennai 600113, India}
\affiliation{Homi Bhabha National Institute, Training School Complex, Anushaktinagar, Mumbai 400 094, India}
\author{Shakti N. Menon}
\affiliation{The Institute of Mathematical Sciences, CIT Campus, Taramani, Chennai 600113, India}
\author{Sitabhra Sinha}
\affiliation{The Institute of Mathematical Sciences, CIT Campus, Taramani, Chennai 600113, India}
\affiliation{Homi Bhabha National Institute, Training School Complex, Anushaktinagar, Mumbai 400 094, India}
\date{\today}
\begin{abstract}
We show that when cells communicate by contact-mediated interactions, heterogeneity in cell shapes and sizes leads to qualitatively distinct collective behavior in the tissue. For inter-cellular coupling that implements lateral inhibition, such disorder-driven transitions can substantially alter the asymptotic pattern of differentiated cells by modulating their fate choice through changes in the neighborhood geometry. In addition, when contact-induced signals influence inherent cellular oscillations, disorder leads to the emergence of functionally relevant partially-ordered dynamical states.
\end{abstract}


\maketitle


Many natural systems, ranging from granular materials to biological tissues and dense crowds,
are characterized by
varying levels of heterogeneity in their structural attributes~\cite{Jaeger1992,Jaeger1996,Gibson2009,Trepat2018,Bottinelli2016,Kulkarni2019}.
This disorder arises via self-organization as a result of interactions between
their numerous constituent units, causing their arrangement to deviate from regular lattice ordering~\cite{Lecuit2007b,Blanchard2009,Julicher2017,Molnar2021}.
A striking example in the biological context is provided
by confluent epithelial tissue, whose
constituent cells are packed together in a high state of disorder, as characterized
by quantitative measures that incorporate the area, perimeter or number of neighbors
of each cell~\cite{Zallen2004,Hilgenfeldt2008,Ragkousi2014}.
Moreover, as the cells communicate with each other, e.g., via the ubiquitous Notch pathway in
which signaling occurs via receptor-ligand binding~\cite{Artavanis1995,Kopan2009,Sprinzak2011},
disorder may also have remarkable functional consequences.
Note that the Notch pathway
effectively implements \textit{lateral inhibition} through which the induction of a specific fate in a
particular cell prevents its immediate neighbors from expressing the same fate~\cite{Barad2010,Sprinzak2010}.
As this is one of the principal mechanisms through which patterning arises in tissues~\cite{Meinhardt1974,Wolpert2011},
any disorder in the geometry of neighborhood contacts that alter
the nature of interactions between adjacent cells consequently affects their fates~\cite{Kuyyamudi2021_c}.
A natural question in this context relates to the relative roles of  local, contact-mediated interactions
and global forces that alter the degree of disorder in shaping the collective behavior of cellular assemblies.

\begin{figure}[tbp!]
\centering
\includegraphics[width=0.99\columnwidth]{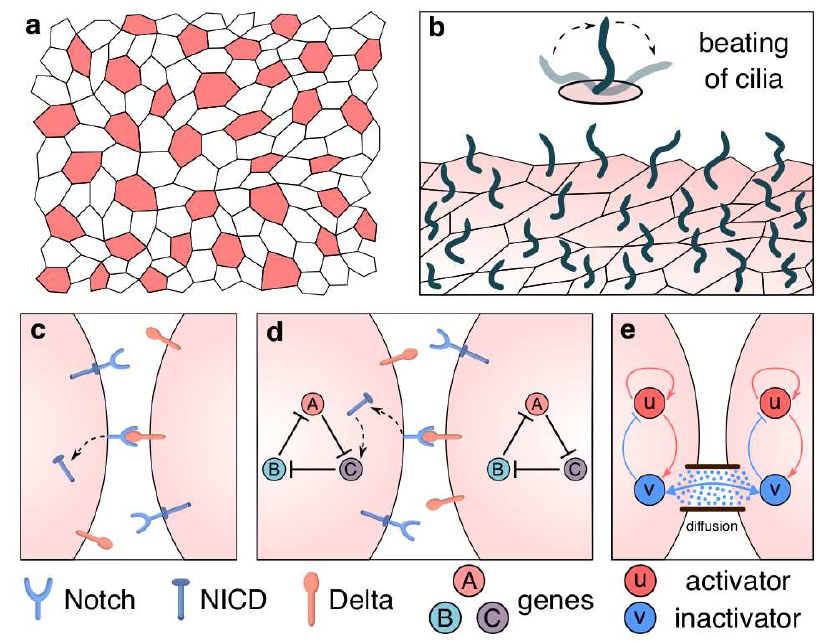}
\caption{\textbf{Communication between neighboring cells in a close-packed disordered
configuration underlie a range of collective behavior.}
(a) Schematic diagram of the spatial arrangement of hair cells (shown in red)
surrounded by supporting cells
in the avian basilar papilla, at an early stage of development~\cite{Goodyear1997}.
(b) Ventral tissue of the marine
animal \textit{Trichoplax adhaerens} illustrated  schematically to show the arrangement of
monociliated epithelial cells~\cite{Ruthmann1986,Smith2014,Bull2021}. Each of the cilia
engage in periodic motion (``beating'', see inset) that helps propel the organism across a surface~\cite{Grell1991,Smith2015}.
(c-e) The key qualitative features of the collective dynamics in such systems 
are seen to be invariant despite differences in the means by which cells communicate and
the dynamics within each cell, e.g., in cells coupled via
\textit{trans-}activation of Notch receptors by Delta ligands resulting in release of a downstream
effector (NICD) [shown in (c)], repressilators coupled by Notch-Delta
signaling (d) and relaxation oscillators coupled via diffusion of the inactivation variable through
inter-cellular bridges such as gap junctions (e).}
\label{fig:fig1}
\end{figure}

A striking illustration of such interplay between disorder and interactions can be seen during
the appearance of a characteristic spatial pattern of cells in the basal papilla (the auditory sensory
organ in all amniotes~\cite{Fritzsch2013}), comprising specialized sensory ``hair cells''
that are separated from
each other by intervening support cells [Fig.~\ref{fig:fig1}~(a)].
As either cell type can arise from the same progenitor cell, the specific fate induced
in a particular cell depends on the cues it receives from its neighborhood~\cite{Goodyear1997}.
In particular, hair cells inhibit their immediate neighbors from
adopting the same fate~\cite{Simpson1990,Barad2010}.
Disorder in the structural arrangement of a cell's
neighborhood can drastically affect these cues and consequently, the resulting fate choice.
More generally, one can investigate novel qualitative features 
in the collective behavior,
such as partially ordered
or ``chimera'' states~\cite{Abrams2004,Singh2011},
that may result from structural heterogeneities.
%
This is particularly relevant where heterogeneity arises through flexibility in cell shapes, typically
observed at the embryonic stage~\cite{Kim2021} but retained
lifelong in simpler animals such as \textit{Trichoplax adhaerens}~\cite{Prakash2021}.
The resulting disordered arrangement of cells in this organism, when coupled to the oscillatory
dynamics of the cilia of each cell, can affect organism-level behavior such as gliding locomotion
along surfaces propelled by collective beating of the
cilia~\cite{Grell1991,Smith2015} [Fig.~\ref{fig:fig1}~(b)].
These examples suggest that the composition and function of tissues
can be altered significantly with increasing heterogeneity in cell sizes and shapes.

In this paper we explicitly demonstrate such transitions with increasing disorder in the arrangement of cells that interact via contact-induced signaling. 
When the interactions between cells implement lateral inhibition, it can influence fate induction
to alter the relative proportions of distinct cell types, and consequently affect development.
We also demonstrate that in tissues where cells are susceptible to random failures in their ability to communicate with neighbors, heterogeneity in the cellular packing geometry makes 
the asymptotic pattern of differentiated cells more robust.
Further, if the inter-cellular interactions modulate activity in the cells, such as
oscillations in molecular concentrations~\cite{Buzsaki2004,Lenz2011,PotvinTrottier2016,Guzzo2018},
we observe that disorder promotes the emergence of the complex spatio-temporal phenomenon
of chimera states.
These are characterized by the coexistence of oscillating cells with those whose activity
has been arrested, and we show that they arise irrespective of whether adjacent cells are coupled through
receptor-ligand binding or by diffusion across bridges such as gap junctions  [Fig.~\ref{fig:fig1}~(c-e)].
Thus, selective deformation of a cellular assembly can drive transitions between dynamical
states marked by different proportions of oscillating elements,
suggesting an intriguing locomotory mechanism in simple multicellular organisms.

\begin{figure}[tbp!]
\centering
\includegraphics[width=0.99\columnwidth]{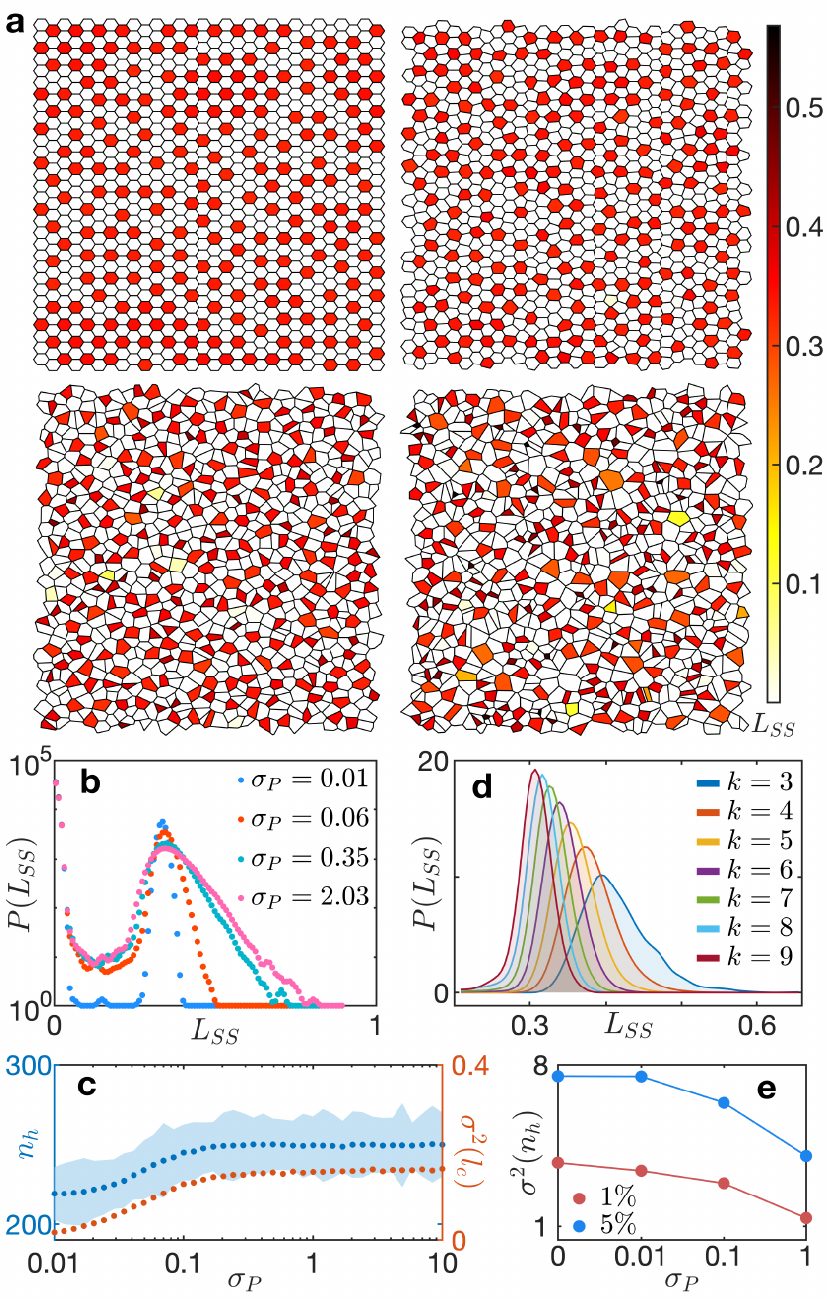}
\caption{\textbf{Higher disorder in the cellular packing configuration allows
a more equitable and robust distribution of cell fates.}
(a) Spatial patterns formed by the steady state Delta ligand concentration, $L_{SS}$ (see color
bar) in an assembly of $N (=900)$ cells resulting from lateral inhibition. The panels represent
increasingly disordered configurations as indicated by the dispersion
of the deviations in cell positions
from those in the regular hexagonal lattice:
$\sigma_P=0$ (top left), $0.01$ (top right), $0.1$ (bottom left) and $1$ (bottom right).
As seen from the bimodal distribution in (b), cells either have very low or high values of $L_{SS}$.
The increase in the width of the high $L_{SS}$ peak with disorder is quantified in (c) which shows that the number of cells $n_h$ (blue dots) in this state increases with the disorder~\cite{note2}.
The shaded region represents the dispersion in $n_h$.
The variance of the cell perimeters ($\sigma^2 (l_c)$, red dots) also rises with disorder
in a qualitatively similar manner.
(d) The asymptotic ligand concentration in a cell appears to be correlated with the number
of its neighbors $k$. The $L_{SS}$ distributions in cells with a specific $k$
monotonically shift to left with increasing $k$, suggesting that cells in states characterized
by higher $L_{SS}$ have fewer neighbors than average.
(e) Greater robustness to damage in the cellular array is seen with increased disorder,
as evident from the reduced variability of fate distribution (measured in terms of dispersion
in $n_h$) with rise in $\sigma_P$ when $1\%$ (red) or $5\%$ (blue) of
randomly chosen cells are rendered inert.
}
\label{fig:fig2}
\end{figure}

To generate disordered cellular configurations
we use the method of Voronoi tessellations to construct two-dimensional space-filling
tilings with non-overlapping polygons that are characterized by varying levels of heterogeneity.
We begin with a regular hexagonal lattice that is then disordered by adding Gaussian noise
$\mathcal{N}(0,\sigma_P)$ to randomly displace each of the generating points or seeds
(initially, the centroids of the hexagons).
The standard deviation $\sigma_P$ can be tuned to yield different levels of heterogeneity.
The extent of disorder in the lattice, measured by the variance
of the perimeters of the cellular polygons $\sigma^2(l_e)$, reaches its maximal value for $\sigma_P \sim 1$ and does not change appreciably on increasing $\sigma_P$ further [Fig.~\ref{fig:fig2}]. 
The strength of coupling between a pair of adjacent cells is assumed to be proportional
to the total length of their interface.
A weighted adjacency matrix \textbf{A}, with $A_{ij}$ representing the overlap between the cells $i$ and $j$, thus provides the information
required to assign interaction strengths between each pair of cells.

We consider contact-induced signaling via Notch receptors located
on the surface of a cell binding to ligands (e.g., Delta) embedded on the membrane
of a neighboring cell (i.e., \textit{trans} binding). This is represented by the following set of equations describing the
time-evolutions of the concentrations of the receptor ($R$), ligand ($L$) and the
Notch intra-cellular domain or NICD ($S$), the downstream effector of the Notch
signaling pathway:
\begin{eqnarray}
\label{eqnR}
  \frac{dR_i}{dt} &=& \beta_R - \gamma_R R_i - k_{cis} R_i L_i - k_{tr} R_i L^{tr}_i,\\
\label{eqnL}
  \frac{dL_i}{dt} &=& \frac{\beta_L K_{s}^h}{K_{s}^h + S^{h}_i}- \gamma_L L_i - k_{cis} R_i L_i - k_{tr} L_i R^{tr}_i,\\
\label{eqnS}
  \frac{dS_i}{dt} &=& k_{tr} R_i L^{tr}_i - \gamma_S S_i.
\end{eqnarray}
Here $R_i^{tr} = \sum_j A_{ij} R_j$ and $L_i^{tr} = \sum_j A_{ij} L_j$ are the weighted sums of receptor and ligand concentrations, respectively,
in the neighborhood of the $i^{th}$ cell.
Earlier studies have shown that lateral inhibition requires strong inhibition of Notch receptors via
\textit{cis} binding (i.e., to ligands on the same cell)~\cite{Barad2010, Sprinzak2010}.
Consistent with this, we choose
$k_{tr}=0.13$ and $k_{cis}=4.64$, which are related to the rates of \textit{trans} activation
and \textit{cis} inhibition, respectively.
The maximal production rates of both
receptors ($\beta_R$) and ligands ($\beta_D$) are chosen to be $100$. The contact-induced
signal is assumed to have a relatively longer life-time so that the degradation rates of the
receptors ($\gamma_R$), ligands ($\gamma_D$) and NICD ($\gamma_S$) are chosen as
$1$, $1$ and $0.1$, respectively. The repression of ligand production by the downstream effector of
Notch signaling pathway is modeled by a Hill function, parameterized by $K_s (=10)$ and $h (=4)$.
The initial concentrations for the ligands and receptors are chosen from an uniform random distribution defined over the domain $[0,10]$.

In presence of strong \textit{cis} inhibition,
only those cells in which ligands far outnumber receptors can engage in
\textit{trans} activation of Notch receptors of neighboring cells. Consequently, the production of
ligands in these cells is inhibited [see Eqn.~(\ref{eqnL})].
The resulting unequal distribution of ligands among cells results in each of them
eventually becoming either
(i) a \textit{receiver} cell having receptors but no ligands, such that it
can only ``receive'' contact-induced inter-cellular signals, or (ii) a \textit{transmitter} cell, which possess ligands but no receptors, such that it can only ``send out" signals.
Mutual competition for \textit{trans}-binding between neighboring cells having more ligands than receptors
is reinforced by the
suppression of ligand production in the cell whose receptors are activated.
Thus, each cell which develops into a transmitter would be surrounded exclusively by cells
which adopt the fate of receivers~\cite{Barad2010}.
This mutual ``repulsion'' between transmitter cells imposes a strong constraint on their numbers
as such cells need to be separated from each other by receiver cells. For example, this requirement
would allow only $\sim N/3$ transmitters in a hexagonal lattice comprising $N$ cells.
However, instead of a regular lattice, if we consider a disordered arrangement, e.g., a sheet of epithelial cells, the total number of transmitter cells allowed in the resulting packing increases noticeably~\cite{note1}.
%
This can be observed from Fig.~\ref{fig:fig2}~(a) which shows the spatial patterns of cellular ligand concentration $L_{SS}$ in the steady state 
as $\sigma_P$ is increased.

The observed bimodal nature of the $L_{SS}$ distribution is invariant to disorder [Fig.~\ref{fig:fig2}~(b)]. Such a distribution allows
a natural segregation of the cells into receivers and transmitters, corresponding to
populations around its lower and higher peaks, respectively.
Further, we note that the cell-cell interface lengths, which crucially
dictate the magnitude of the contact-induced signal, exhibit higher variance with increased disorder [Fig.~\ref{fig:fig2}~(c)]. 
This
is mirrored in the rise of the
number of transmitter cells $n_h$ with $\sigma_P$
[Fig.~\ref{fig:fig2}~(c)].
The broadening of the peaks in the $L_{SS}$ distribution with increasing heterogeneity
of the cellular configuration can be understood in terms of the role that the degree $k$
of a transmitter cell (i.e., the number of cells in its immediate neighborhood) plays in
determining the steady state ligand concentration.
Fig.~\ref{fig:fig2}~(d) shows that the ligand distribution of cells
having exactly $k$ neighbors shifts to the right with decreasing $k$. Thus, the peak-broadening
with $\sigma_P$ [Fig.~\ref{fig:fig2}~(b)] can be attributed to a higher density of
transmitter cells with lower $k$ (compared to the regular lattice). 
With increasing heterogeneity,
transmitter cells have
fewer neighbors on average, implying that more
cells can become transmitters as their number is only limited by the
constraint that no two of them can be neighbors.

As transmitters and receivers correspond to cells with distinct fates,
the change in the relative proportion of such cells resulting from disordered cellular arrangements 
suggests that this can alter the course of development.
Heterogeneity also makes the spatial pattern robust against damage that may strike a cell at random,
disabling it from taking part in inter-cellular signaling~\cite{SI}.
%
This is quantified by the dispersion in $n_h$, the number of cells likely to become transmitters,
shown in Fig.~\ref{fig:fig2}~(e) for two different fractions of randomly damaged cells.
As $\sigma_P$ is increased, the variance decreases noticeably, suggesting that more
disordered cellular configurations have less variability in terms of the relative proportion of cells
having distinct fates.
\begin{figure}[tbp!]
\centering
\includegraphics[width=0.99\columnwidth]{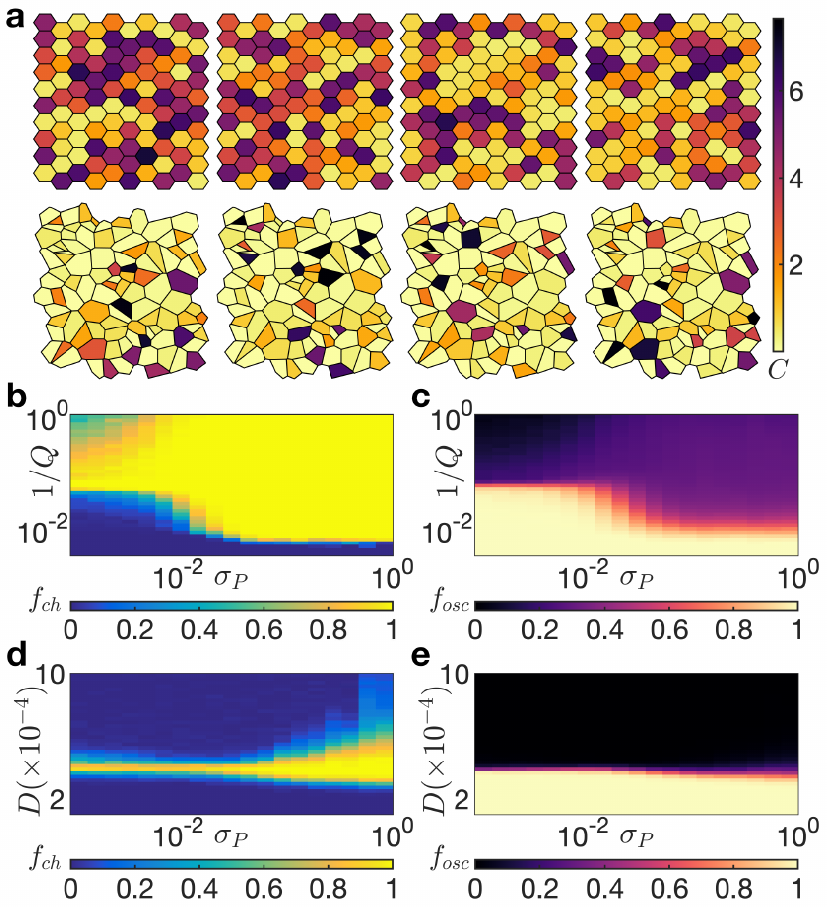}
\caption{\textbf{Disorder promotes the coexistence of qualitatively distinct behaviors
(chimera states)
in the collective dynamics
of cellular oscillators coupled via contact-mediated interactions.}
(a) Instantaneous states of oscillator arrays that are (top row) ordered ($\sigma_P=0$), or (bottom row) disordered
to the maximum extent ($\sigma_P=1$), shown at times separated by an interval
that is $1/4$ of the
oscillation period of an uncoupled cell.
Colors represent the expression level of one of the genes ($C$) comprising the
oscillating repressilator circuit~\cite{movie}.
(b-c) The fraction of realizations $f_{ch}$ in which chimera states are observed (b) and the
mean fraction of cells that continue to oscillate $f_{osc}$ (c) shown as a function of the
disorder in cellular arrangement ($\sigma_P$), as well as the 
strength of inter-cellular interaction induced repression (measured as $1/Q$).
(d-e) Qualitatively similar behaviors in (d) $f_{ch}$ and (e) $f_{osc}$ are shown by systems of diffusively coupled relaxation oscillators. Increasing the diffusion constant $D$ beyond a critical
value leads to cessation of activity through oscillator death.
In both systems, increased disorder allows configurations with coexisting oscillating and non-oscillating cells to exist over a much larger range. 
}
\label{fig:fig3}
\end{figure}

The model system reported above focuses only on signaling between cells, without considering
how such signals can alter the intra-cellular dynamics. 
However, Notch signaling is known to play an important role in processes such as somitogenesis~\cite{Lewis2003,Kuyyamudi2021_b}
and tissue growth by cell division~\cite{Deng2001,Shimojo2008}, where it takes part in
non-trivial dynamics, involving periodically varying molecular concentrations.
Therefore, we now consider cellular dynamics described by an oscillating circuit comprising three cyclically repressing genes 
$A$, $B$ and $C$~\cite{Elowitz2000}, any one of which is assumed to be regulated by the inter-cellular signal $S$. 
The collective dynamics of these cellular oscillators coupled by Notch signaling (specifically, by $S$ inhibiting
$C$)
can be described by the time-evolutions
of $R$, $L$ and $S$ described earlier (with $h=2$, other parameter values unchanged), augmented by
the following equations for the gene
products:
\begin{eqnarray*}
  \frac{dA_i}{dt} &=& \alpha\left[\frac{K^g}{K^g + C_{i}^g}\right] - \frac{A_i}{\tau} ,\\
  \frac{dB_i}{dt} &=& \alpha\left[\frac{K^g}{K^g + A_{i}^g}\right] - \frac{B_i}{\tau} ,\\
  \frac{dC_i}{dt} &=& \alpha\left[\frac{K^g}{K^g + B_{i}^g}\right]\left[ \frac{Q^g}{Q^g + S_{i}^g}\right] - \frac{C_i}{\tau}.
\end{eqnarray*}
The maximal production rates $\alpha (=10)$, mean lifetimes $\tau (=1)$, and
the parameters $K (=1)$ and $g(=4)$ of the Hill functions describing the cyclic repression
are chosen to ensure oscillations in absence of inter-cellular coupling.
The repression of gene expression by 
$S$ is also modeled by a Hill function, parameterized by the exponent $g(=4)$ and the
half-saturation constant $Q$. Upon strengthening the repression (i.e., increasing $1/Q$), the
collective dynamics shows a transition from global oscillations to a quiescent state
that arises from oscillation arrest.
Introducing disorder in the cellular arrangement leads to the emergence of chimera states
in the collective dynamics [Fig.~\ref{fig:fig3}~(a)].
Depending on the context, such states have diverse implications, e.g., in growing tissues
characterized by coupled cell-cycle oscillations, they potentially contribute to morphogenesis by
selective growth, as
only the cells that continue to oscillate can keep dividing~\cite{Kuyyamudi2021_c}.
Chimera states could also shape
the trajectory of an organism whose locomotion is guided by oscillatory beating of ciliary rotors [as
in \textit{T. adhaerens}, Fig.~\ref{fig:fig1}~(b)] by selectively rendering certain cilia immotile.
As heterogeneity is increased, the range of coupling strengths for
which chimera states can be observed increases markedly, appearing even for very weak interactions
between cells [Fig.~\ref{fig:fig3}~(b)].
This is accompanied by cessation of oscillations in the majority of the cells even at low levels
of repression [$Q \sim 10^{2}$, see Fig.~\ref{fig:fig3}~(c)]. 

The generality of these results can be demonstrated by using a generic description of relaxation oscillations to describe the dynamical behavior of each cell. 
This involves a fast activation component $u$ and
a relatively slower inactivation (or inhibitory) variable $v$, whose time-evolution is given by the Fitzhugh-Nagumo equations~\cite{Fitzhugh1961,Nagumo1962,Sinha2014}. 
The lateral inhibition resulting from the receptor-ligand binding mediated interaction mechanism
is implemented by diffusive coupling via $v$ between the
oscillators~\cite{Singh2013,Janaki2019}, viz., $du_i/dt = u_i(1 - u_i)(u_i - \phi) - v_i$, $dv_i/dt
= \epsilon (\kappa u - v - b) + D\Sigma_j A_{ij}(v_j - v_i)$, where ${\mathbf A}$ is the weighted
adjacency matrix.
The parameters $\phi (= 0.139)$, $b (=0.17)$ and $\kappa (=0.6)$ specify the kinetics,
and $\epsilon (=0.001)$ is the recovery rate.
The strength of diffusive coupling $D$ between neighboring oscillators, which is the analog of the
parameter $1/Q$ for the system of coupled repressilators,
is altered systematically in our simulations. 
Note that, increasing disorder in the cellular arrangement alters the diffusive flux between coupled cells,
which is proportional to the length of the corresponding interface. 
This is consistent with the linear extent of the cellular interface being proportional to 
the density of gap junctions (or other structures that
bridge the cytoplasms of cells),
provided that they are homogeneously distributed across the cell membrane.
Fig.~\ref{fig:fig3}~(d) shows that, as in the case of Notch coupled repressilators, increasing
heterogeneity promotes the existence of chimera states over a range of $D$~\cite{SI}.
They can be characterized by the fraction of oscillating cells $f_{osc}$ lying between
$0$ and $1$, with the chimera region straddling the boundary
separating global synchronization ($f_{osc} = 1$)
from complete quiescence ($f_{osc} = 0$) [Fig.~\ref{fig:fig3}~(e)].
We have
verified that the qualitative features of the transition remain invariant
to stochastic fluctuations in molecular concentrations~\cite{SI}.
Thus, disorder-driven transitions 
appear to be a general phenomenon that might be observed in systems with
different mechanisms for oscillations and diverse types of inter-cellular interactions.

%

To conclude, 
we have shown that changes in the packing arrangement of cells, as they
become more heterogeneous, modulate their collective behavior arising
from inter-cellular interactions implementing lateral inhibition.
This can play a key role in determining the relative proportions of specialized cells, such as
those expressing thoracic bristles in \textit{Drosophila}~\cite{Simpson1990,Heitzler1991} or
neurons that arise by selective differentiation of progenitor cells in the mammalian nervous system~\cite{Kageyama2008}.
Furthermore, disorder contributes to the robustness of the specific composition of tissues
to cell damage.
The promotion of chimera states in the collective dynamics of cells upon increasing their heterogeneity
has multiple implications, including providing a mechanism for selective regulation of growth
in confluent tissue or establishing left-right asymmetry by altering large-scale patterns in
ciliary motion during development~\cite{Nonaka1998}.
Our results can be experimentally corroborated in epithelial sheets of cells interacting via contact-mediated
coupling and characterized by varying degrees of disorder. For example, \textit{T. adhaerens} whose cells are capable of continually, and radically, altering their shape~\cite{Pearse2007,Prakash2021}, can provide a test-bed for relating disordered configurations of epithelial
tissue with the collective motion of the cilia attached to every cell.
Biofilms comprising oscillating bacterial cells that coordinate their activity
by electrical signaling can be another potential experimental system to
explore how disordered arrangements alter collective dynamics~\cite{Prindle2015,Humphries2017}.
Further work may also elucidate the potential
role of disorder, which arises naturally via cellular remodeling during development,
in shaping morphogenesis.

\begin{acknowledgments}
SNM has been supported by the IMSc Complex Systems Project (12th
Plan), and the Center of Excellence in Complex Systems and Data
Science, both funded by the Department of Atomic Energy, Government of
India. The simulations required for this work were
performed using the IMSc High Performance Computing facility (hpc.imsc.res.in) [Nandadevi].
\end{acknowledgments}


\begin{thebibliography}{99}
\bibitem{note2}
We count $n_h$ as the number of cells having $L_{SS} >0.2$. We have verified
that the specific choice of the threshold value does not qualitatively change
our results.
\end{thebibliography}
%

\clearpage
\onecolumngrid

\setcounter{figure}{0}
\renewcommand\thefigure{S\arabic{figure}}
\renewcommand\thetable{S\arabic{table}}

\vspace{1cm}
\begin{center}
\textbf{\large{SUPPLEMENTARY INFORMATION}}\\

\vspace{0.5cm}
\textbf{\large{Contact-mediated signaling enables disorder-driven transitions in cellular assemblies}}\\
\vspace{0.5cm}
\textbf{Chandrashekar Kuyyamudi, Shakti N. Menon and Sitabhra Sinha}
\end{center}
\section*{List of Supplementary Figures and Movies}
\begin{enumerate}
\item Fig S1: Disordered cellular arrangement can promote cell fate choices that are
constrained by lateral inhibition.
\item Fig S2: Disorder enhances fate pattern robustness against random
cell damage.
\item Fig S4: Reduced variability of fate distribution with increased disorder in tissues
subject to random cell damage.
\item Fig S5: Disorder promotes emergence of chimera states in diffusively coupled
relaxation
oscillators.
\item Fig S4: Chimera states are robust with respect to noise.
\item Movie S1: Time-evolution of $C$ gene expression in an ordered array of repressilators
interacting by contact-induced signaling.
\item Movie S2: Time-evolution of $C$ gene expression in a disordered arrangement of repressilators
interacting by contact-induced signaling.
\item Movie S3: Time-evolution of the inactivation variable $y$ in an ordered array of diffusively
coupled relaxation oscillators.
\item Movie S4: Time-evolution of the inactivation variable $y$ in a disordered arrangement of
diffusively
coupled relaxation oscillators.
\end{enumerate}

\section{Numerical details}
The dynamics of each cell in the various models investigated here are described by systems of  coupled ordinary differential equations (ODEs). We solve these coupled ODEs using an adaptive numerical integrator (implemented in the differential equations module in Julia programming language, ver. 1.6.1). The initial values of the dynamical variables have been chosen from
uniform random distributions.

The steady state distributions of ligand concentrations shown in Fig.~2~(b) in the
main text have been obtained by Gaussian kernel smoothing, sampling over $300$ independent trials for four different values of $\sigma_P$.

In Fig.~2~(e) shown in the main text, we consider $10$ different lattices and $30$ different choices of damaged cells for each level of disorder ($\sigma_P$) and measure $n_h$ for each of the trials. We quantify the robustness with respect to random cell damage by measuring the variance in $n_h$ across $300$ different realizations. From the distributions obtained after Gaussian kernel smoothing shown in Fig.~S3, we observe that while the mean value of $n_h$ increases with $\sigma_P$,
the variance [$\sigma^2(n_h)$] noticeably decreases with increase in the level of disorder in the lattice.

\begin{figure}[ht]
\includegraphics[width=\textwidth]{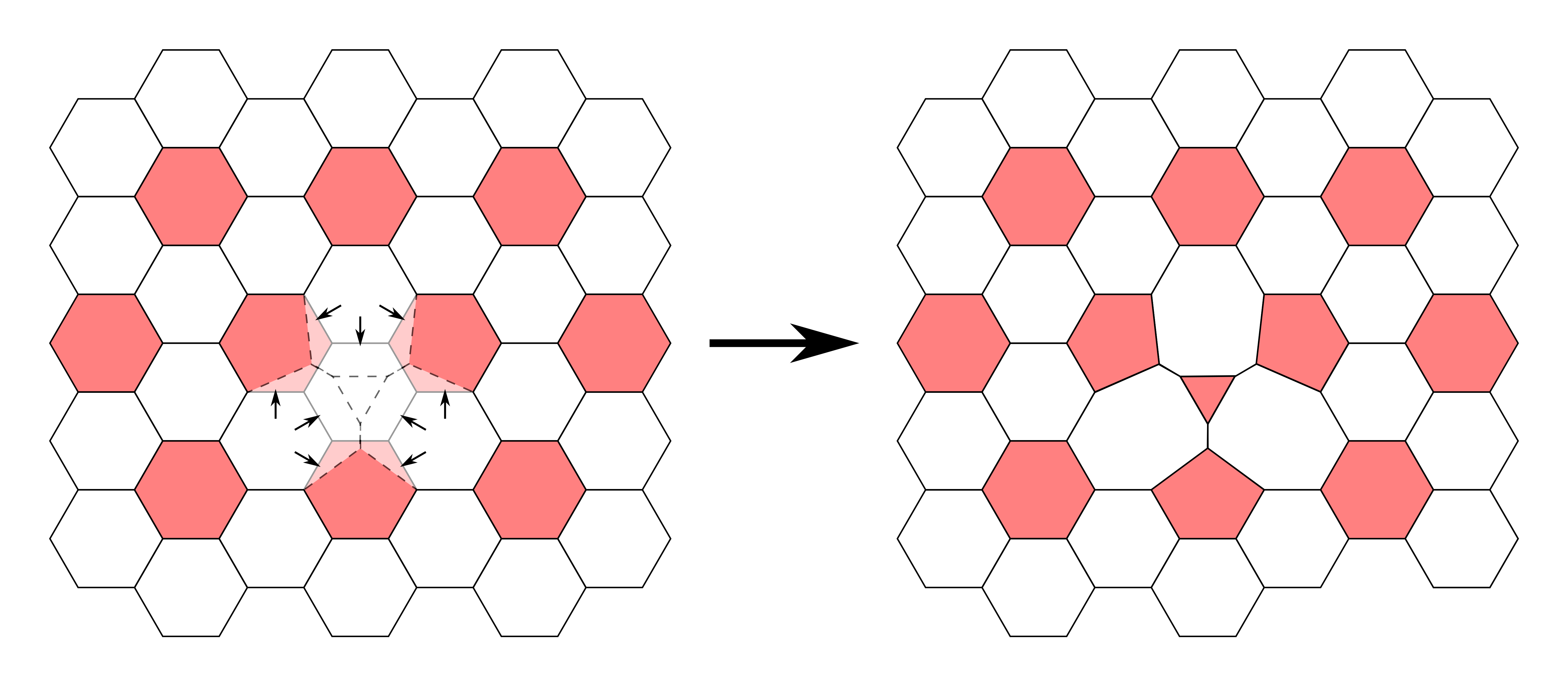}
\caption{\textbf{Disordered cellular arrangement can promote cell fate choices that are
constrained by lateral inhibition.} The schematic diagrams represent a completely ordered
hexagonal close-packed lattice of cells (left) that is deformed to yield the locally disordered
arrangement comprising polygons that have either more or fewer than six neighbors (right).
Colored polygons represent cells that have adopted a specialized fate, e.g., transmitter cells in the example discussed in the main text. This fate choice is assumed to be subject to lateral
inhibition, so that adjacent cells cannot both be colored.
Note that a hexagon that was in contact with three colored polygons in the ordered case
becomes triangular upon deformation and is free to be colored (i.e., adopt the specialized cell fate)
as it no longer neighbors any other colored cell. Thus, the number of colored cells increase by $1$,
even though the number of cells and the total number of cell-cell contacts are conserved.}
\label{figS5}
\end{figure}

\begin{figure}[ht]
\includegraphics[width=\textwidth]{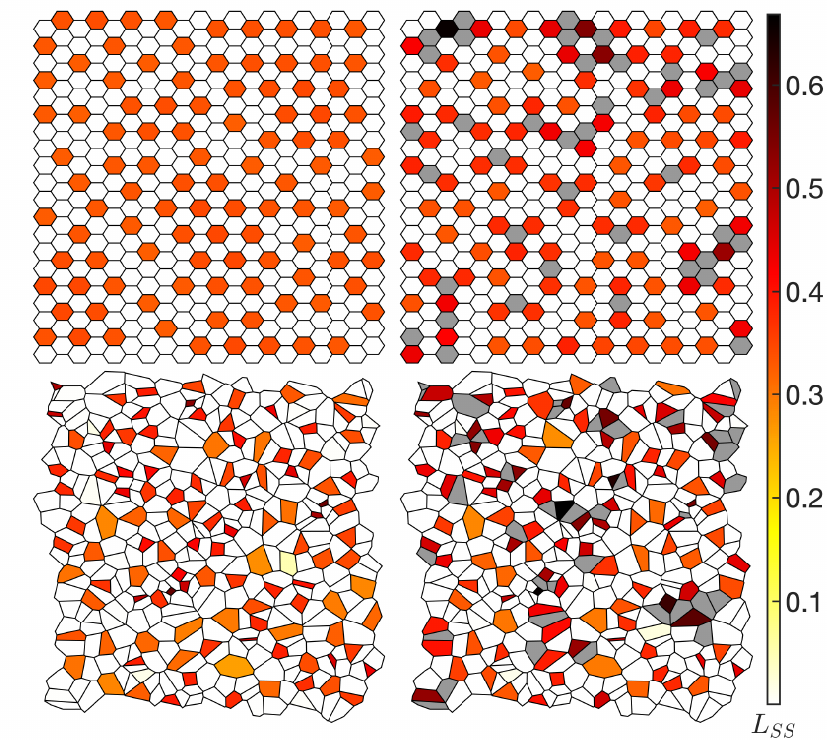}
\caption{\textbf{Disorder enhances fate pattern robustness against random
cell damage.} The steady state spatial arrangement of cell fates
in (top row) the absence of disorder ($\sigma_P = 0$) and (bottom row) with maximal disorder ($\sigma_P = 1$), comparing
the situation (left) where all cells are signaling with the case (right) in which $10\%$ of the
cells (randomly chosen) are inert. The colors
represent the steady state values of the Delta ligand concentration, $L_{SS}$.
The change in the fraction of cells attaining minority fate as a consequence of disruption
resulting in loss of signaling ability by part of the population, is seen to be less when the
tissue is more disordered.}
\label{figS1}
\end{figure}

\begin{figure}[ht]
\includegraphics[width=\textwidth]{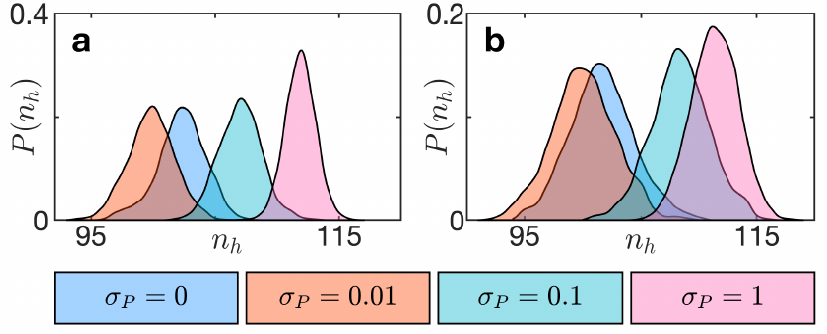}
\caption{\textbf{Reduced variability of fate distribution with increased disorder in tissues
subject to random cell damage.}
The distributions of $n_h$ are shown for different $\sigma_P$ (see key) when (a) $1\%$ and (b) $5\%$ of
randomly chosen cells are rendered inert.
We note that although
the mean increases, the dispersion in $n_h$ decreases with increasing disorder,
as indicated explicitly in Fig.~2~(e) in the main text. }
\label{figS2}
\end{figure}

\begin{figure}[ht]
\includegraphics[width=\textwidth]{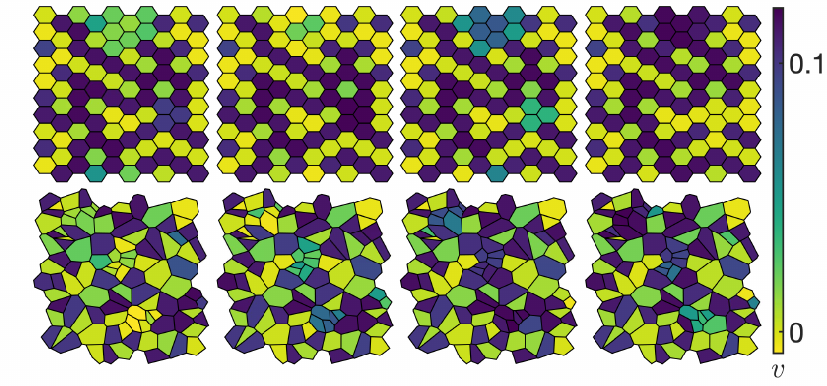}
\caption{\textbf{Disorder promotes emergence of chimera states in diffusively coupled
relaxation
oscillators.} Snapshots of (top row) an ordered lattice ($\sigma_P = 0$) and (bottom row) a maximally disordered lattice
($\sigma_P=1$) comprising $N=100$ relaxation oscillators
shown at different times that are separated by an interval
$\tau_p/4$ where $\tau_p$ is the
oscillation period of an uncoupled cell.
The colors represent the instantaneous value of the inactivation variable $y$ in the
Fitzhugh-Nagumo model used to describe the relaxation oscillator dynamics.
Movies included in the Supplementary Information show the time-evolution of the dynamics in the
arrays.}
\label{figS3}
\end{figure}

\begin{figure}[ht]
\includegraphics[width=0.7\textwidth]{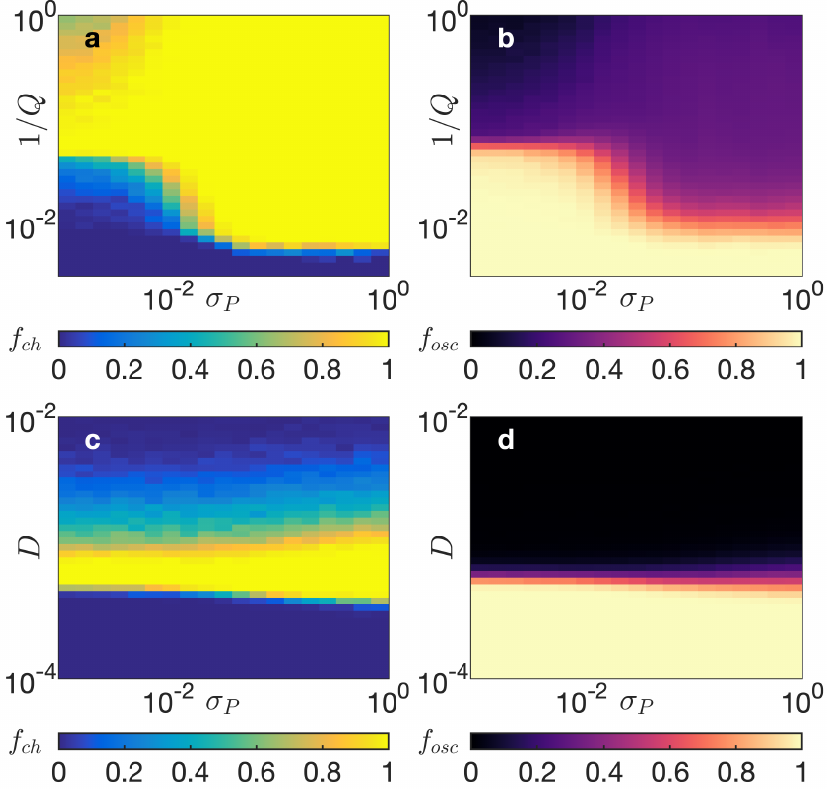}
\caption{\textbf{Chimera states are robust with respect to noise.}
Stochastic perturbation of the intra-cellular dynamics has been implemented by
adding the term $\eta {\bm X} dW$ to the deterministic equations
$d {\bm X} = \mathcal{F}_{\bm X} dt$ shown in the main text (${\bm X}: \{A,B,C\}$
for the repressilator, top row, and ${\bm X}: \{u,v\}$ for the Fitzhugh-Nagumo oscillator, bottom
row). The parameter $\eta$ ($=0.01$ for the results shown here) is the strength of the noise
and $dW$ is a Wiener process.
It is evident that the fraction of realizations $f_{ch}$ in which chimera states are observed (a,c)
and the mean fraction of cells that continue to oscillate $f_{osc}$ (b,d) for the two
systems subject to noise are almost identical to that in the absence
of noise (shown Fig.~3~(b-e)  in main text).
Note that the additive stochastic term induces oscillations in the FHN model, so that
larger values of $D$ are required to
arrest activity compared to the deterministic situation.}
\label{figS4}
\end{figure}

\end{document}